\documentclass{llncs}
\usepackage{makeidx}  

\usepackage{fainekos-macros}

\usepackage{makeidx}  
\usepackage{amsmath} 
\usepackage{amssymb}  
\usepackage{graphicx}
\usepackage{stmaryrd}
\usepackage{tikz}
\usepackage{paralist}
\usepackage{subcaption}
\usetikzlibrary{arrows}
\usepackage{acronym}
\usepackage[obeyFinal]{todonotes}
\usepackage{pgfplots}
\usepackage{ifthen}
\usepackage{hyperref} 
\usepackage{mathtools}
\usepackage{balance}

\acrodef{PDA}{Peak Detection Algorithm}
\acrodef{ICD}{Implantable Cardioverter Defibrillator}
\acrodef{QRE}{Quantitative Regular Expression}
\acrodefplural{QREs}{Quantitative Regular Expressions}
\acrodef{RE}{Regular Expression}
\acrodef{EGM}{electrogram}
\acrodef{VF}{Ventricular Fibrillation}
\acrodef{VT}{Ventricular Tachycardia}

\newcommand{\qre}[1]{\text{\sf{#1}}}
\newcommand{\freeze}{\downarrow}
\newcommand{\CLTLF}{\text{CLTL}^{\freeze}}
\newcommand{\nxt}{X}

\newcommand{\splitop}[1]{split\!-\!#1}

\newcommand{\val}[1]{\ensuremath{\llbracket #1 \rrbracket}}

\newcommand{\egm}{x}

\newcommand{\wav}{W\!_\egm}

\begin{document}
\frontmatter          
\pagestyle{headings}  
\mainmatter              
\title{Quantitative Regular Expressions for Arrhythmia Detection Algorithms}
\titlerunning{Irregular rhythms}  
%
\author{Houssam Abbas\inst{1} \and Alena Rodionova\inst{2}
\and Ezio Bartocci\inst{2}\and \\ Scott A. Smolka\inst{3} \and Radu Grosu\inst{2}}
\authorrunning{Houssam Abbas et al.} 
\institute{
Department of Electrical \& Systems Engineering, University of Pennsylvania, USA\\
\email{habbas@seas.upenn.edu},
\and
Cyber-Physical Systems Group, Technische Universit\"at Wien, Austria\\
\email{\{alena.rodionova,ezio.bartocci,radu.grosu\}@tuwien.ac.at}
\and
Department of Computer Science, Stony Brook University, USA\\
\email{sas@cs.stonybrook.edu}
}

\maketitle              
\begin{abstract}
Motivated by the problem of verifying the correctness of arrhythmia-detection algorithms, we present a formalization of these algorithms in the language of \textit{Quantitative Regular Expressions}. 
QREs are a flexible formal language for specifying complex numerical queries over data streams, with provable runtime and memory consumption guarantees.
The medical-device algorithms of interest include \textit{peak detection} (where a peak in a cardiac signal indicates a heartbeat) and various \textit{discriminators}, each of which uses a feature of the cardiac signal to distinguish fatal from non-fatal arrhythmias.
Expressing these algorithms' desired output in current temporal logics, and implementing them via monitor synthesis, is cumbersome, error-prone, computationally expensive, and sometimes infeasible. 

In contrast, we show that a range of peak detectors (in both the time and wavelet domains) and various discriminators at the heart of today's arrhythmia-detection devices are easily expressible in QREs.
The fact that one formalism (QREs) is used to describe the desired end-to-end operation of an arrhythmia detector opens the way to formal analysis and rigorous testing of these detectors' correctness and performance. 
Such analysis 
could alleviate the regulatory burden on device developers when modifying their algorithms.
The performance of the peak-detection QREs is demonstrated by running them on real patient data, on which they yield results on par with those provided by a cardiologist.
\end{abstract}

\section{Introduction}
\label{sec:introduction}

\begin{figure*}[t!]
	\centering
	\begin{subfigure}[t]{0.5\textwidth}
		\centering
		\includegraphics[height=1.2in]{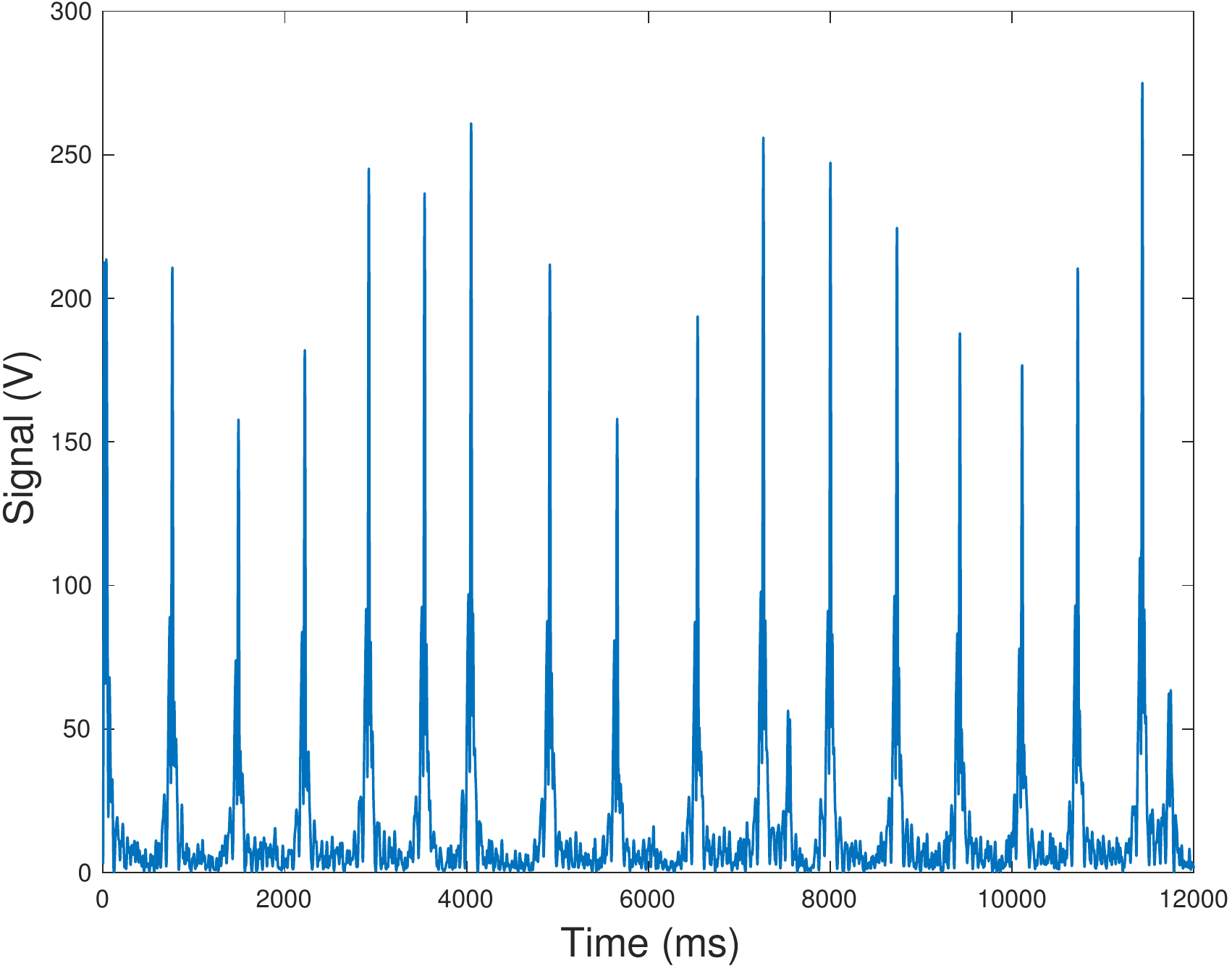}
	\end{subfigure}%
	~ 
	\begin{subfigure}[t]{0.5\textwidth}
		\centering
		\includegraphics[height=1.2in]{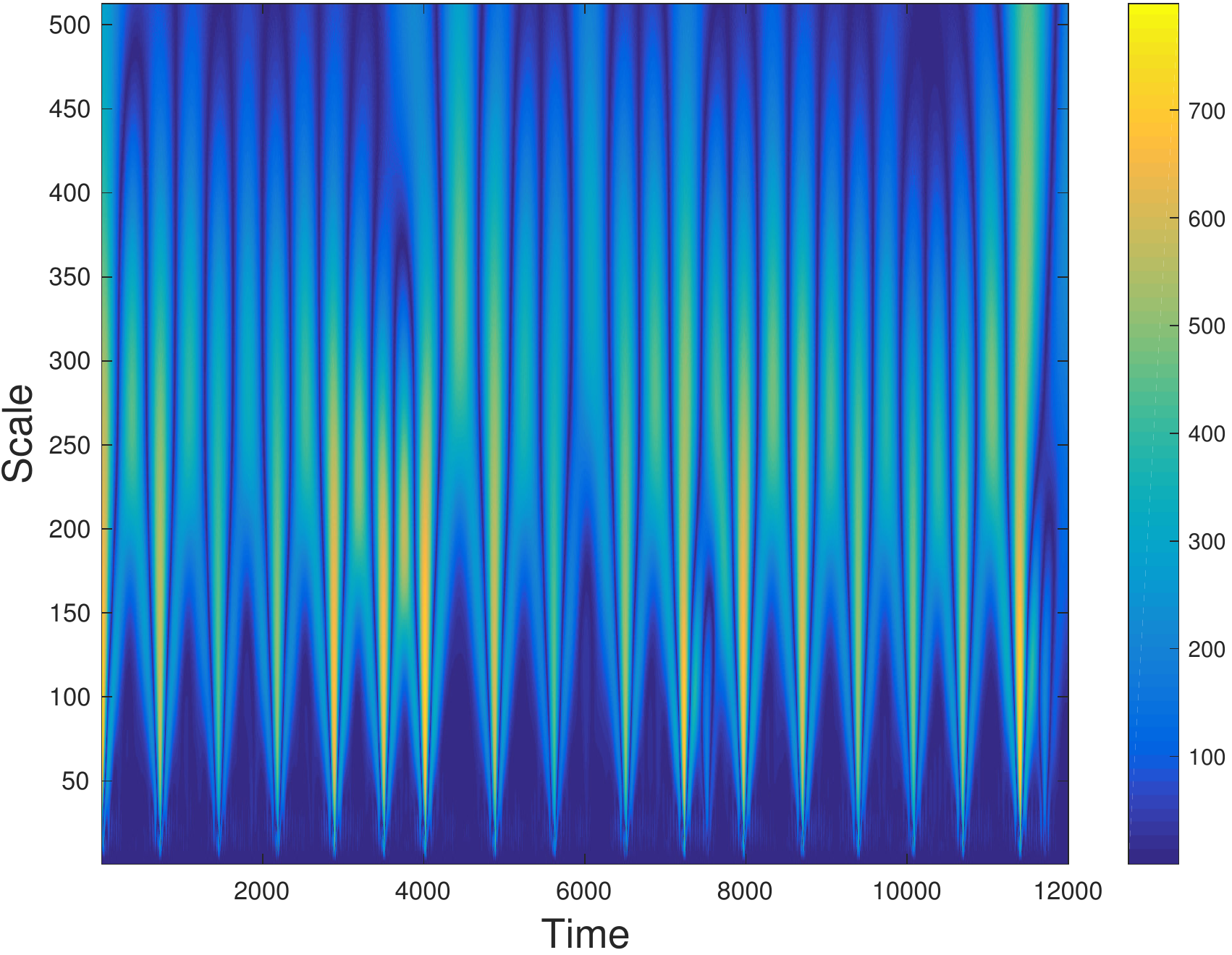}	
	\end{subfigure}
	\caption{\small Rectified EGM during normal rhythm (left) and its CWT spectrogram (right)}
	\label{fig:nsr and its cwt}
\end{figure*}

Medical devices blend signal 
processing (SP) algorithms with decision algorithms such that the performance and correctness of the latter critically depends on that of the former.  
As such, analyzing a device's decision making in isolation of SP offers at best an incomplete picture of the device's overall behavior.  
For example, an \ac{ICD} will 
first perform \textit{Peak Detection} (PD) on its input voltage signal, also known 
as an \textit{electrogram} (see Fig.~\ref{fig:nsr and its cwt}). 
The output of PD is a timed boolean signal where a~1 indicates a peak (local extremum) produced by a heartbeat,
which is used by the downstream \textit{discrimination algorithms} to differentiate between fatal and 
non-fatal rhythms.
Over-sensing (too many false peaks detected) and under-sensing (too many true peaks missed) can be responsible for as much as 10\% of an ICD's erroneous decisions~\cite{Swerdlow14_SensingTroubleshooting}, as they lead to inaccuracies in estimating the heart rate and in calculating important timing relations between the beats of the heart's chambers.

Motivated by the desire to verify ICD algorithms for cardiac 
arrhythmia discrimination, \textit{we seek a unified formalism for expressing and analysing the PD and discrimination tasks commonly found in ICD algorithms.}
A common approach would be to view these tasks as one of checking that the cardiac signal satisfies certain requirements, express these requirements in temporal logic, and obtain the algorithms by monitor synthesis. 
For example, PD evaluates to~1 if the signal (in an observation window) contains a peak, 
while the \textit{V-Rate} discriminator evaluates to~1 if the average heart rate exceeds a certain threshold. 

As discussed in Section~\ref{sec:manylogics}, however, this approach quickly leads to a fracturing of the formalisms: PD algorithms and the various discriminators require different logics, and some simply cannot be expressed succinctly (if at all) in any logic available today.
Thus, despite the increasingly sophisticated variety of temporal logics that have appeared in the literature~\cite{Donze2012,BrimDSV14}, they are inadequate for expressing the operations of PD and discrimination succinctly.
It should be noted that PD is an extremely common signal-processing primitive used in many domains, and forms of discrimination appear in several cardiac devices besides ICDs, such as Implantable Loop Recorders and pacemakers.
Thus the observed limitations of temporal logics extend beyond just ICD algorithms.

PD and discrimination both require reasoning, and performing a wide range of numerical operations, over data streams, where the data stream is the incoming cardiac electrogram observed in real-time.
For example, a commercial peak detector (demonstrated in Section~\ref{sec:experiments}) defines a peak as a value that exceeds a certain time-varying threshold, and the threshold is periodically re-initialized as a percentage of the previous peak's value.
As another example, the \textit{Onset} discriminator compares the average heart rate in two successive windows of fixed size.
Thus, the desired formalism must enable value storage, time freezing, various arithmetic operations, and nested computations, \textit{while remaining legible and succinct, and enabling compilation into efficient implementations}.

We therefore propose the use of \acfp{QRE} to describe 
(three different) peak detectors and a common subset of discriminators.
\acp{QRE}, described in Section~\ref{sec:qre primer}, are a \textit{declarative} formal language based on classical regular expressions for specifying \textit{complex numerical 
queries on data streams}~\cite{Alur2016}.
\acp{QRE}' ability to interleave user-defined computation at any 
nesting level of the underlying regular expression gives them significant expressiveness.
(Formally, \acp{QRE} are equivalent to the streaming composition of regular functions~\cite{AlurFR2014}).
\acp{QRE} can also be compiled into runtime- and memory-efficient implementations, which is an important consideration for implanted medical devices.

To demonstrate the versatility and suitability of \acp{QRE} for our task, we focus on PD in the rest of the paper, since it is a more involved than any single discriminator.
Three different peak detectors are considered (Section~\ref{sec:peaks in wav domain}): 
\begin{inparaenum}
	\item detector WPM, which operates in the wavelet domain,
	\item detector WPB, our own modification of WPM that sacrifices accuracy for runtime, and
	\item detector MDT, which operates in the time domain, and is implemented in an \ac{ICD} on the market today. 
\end{inparaenum}
For all three, a \ac{QRE} description is derived (Section~\ref{sec:qre PD}).
The detectors' operations is illustrated by running them on real patient electrograms (Section~\ref{sec:experiments}).

In summary, our contributions are:
\begin{itemize}
	\item We show that a common set of discriminators is easily encoded as \acp{QRE}, and compare the \acp{QRE} to their encoding in various temporal logics.
	\item We present two peak detectors based on a general wavelet-based characterization of peaks.
	\item We show that the wavelet-based peak detectors, along with a commercial time-domain peak detector found in current ICDs, are easily
and clearly expressible in \acp{QRE}. 
	\item We implement the \acp{QRE} for peak detection and demonstrate their capabilities on real patient data.
\end{itemize}

\section{Challenges in Formalizing ICD Discrimination and Peak Detection}
\label{sec:manylogics}



This section demonstrates the difficulties that arise when using temporal logic to express the discrimination and peak-detection tasks common to all arrhythmia-detection algorithms.
Specifically: different discriminators require the use of different logics, whose expressive powers are not always comparable; 
the formulas quickly become unwieldy and error-prone;
and the complexity of the monitor-synthesis algorithm, \textit{when} it is available, rapidly increases due to nesting of freeze quantification.
On the other hand, it will be shown that \acp{QRE} are well-suited to these challenges: 
all tasks are expressible in the \ac{QRE} formalism,
the resulting expressions are simple direct encodings of the tasks, 
and their monitors are efficient.
The syntax and semantics of the logics will be introduced informally as they are outside the scope of this paper.

An ICD \textit{discriminator} takes in a finite discrete-timed signal $w: \{0,\ldots,T\} \rightarrow D$.
(Signal $w$ will also sometimes be treated as a finite string in $D^*$ without causing confusion).
The discriminator processes the signal $w$ in a sliding-window fashion. 
When the window is centered at time instant $t$, the discriminator computes some feature of the signal (e.g., the average heart rate) and uses this feature to determine if the rhythm displays a potentially fatal arrhythmia in the current window (at time $t$).
The ICD's overall Fatal vs Non-Fatal decision is made by combining the decisions from all discriminators.

In what follows, several discriminators that are found in the devices of major ICD manufacturers are described.
Then for each discriminator, after discussing the challenges that arise in specifying the discriminator in temporal logic, a \ac{QRE} is given that directly implements the discriminator.
This will also serve as a soft introduction to \ac{QRE} syntax.
Fix a data domain $D$ and a cost domain $C$.
For now, we simplify things by viewing a QRE $\qre{f}$ as a regular expression $r$ along with a way to assign costs to strings $w \in D^*$.
If the string $w$ matches the regular expression $r$, then the QRE maps it to $\qre{f}(w) \in C$.
If the string does not match, it is mapped to the undefined value $\bot$.
The QRE's computations can use a fixed but arbitrary set of \textit{operations} (e.g., addition, max, or insertion into a set).
Operations can be thought of as arbitrary pieces of code.

The first example of discriminator checks whether the number of heartbeats in a one-minute time interval is between 120 and 150.
This requires the use of a counting modality like that used in CTMTL~\cite{Krishna16_CountingMTL}. 
If $p$ denotes a heartbeat, then the following CTMTL formula evaluates to true exactly when the number of heartbeats lies in the desired range: $\text{C}_{[0,59]}^{\geq 120}p\,\wedge\,\text{C}_{[0, 59]}^{\leq 150}p$.

This is equally easily expressed as a \ac{QRE}: match $60$ signal samples (at a 1Hz sampling rate), and at every sample where $p$ is true (this is a heartbeat), add 1 to the cost, otherwise add 0. Finally, check if the sum is in the range:
\begin{equation*}
\mathtt{inrange}(iter_{60}\!-\!add(p?1\textrm{ else } 0))
\end{equation*}

The second discriminator determines whether the heart rate increases by at
least 20\% when measured over consecutive and disjoint windows of 4~beats.
In logic, this requires explicit clocks, such as those used in Explicit Clock Temporal Logic XCTL~\cite{Harel90_XCTL}, since the
beat-to-beat delay is variable. 
So let $T$ denote the time state (which keeps track of time) and let the $x_i$'s be rigid clock variables that store the times at which $p$ becomes true.
The following XCTL formula expresses the desired discriminator:
\begin{align*}
\always(p\,\wedge (x_1=T)\,\wedge\, \eventually(
p\,\wedge \ldots \eventually
(p\,\wedge\, (x_9=T)\,
\wedge\,[(x_5-x_1)\cdot 0.8 \geq x_9-x_5] )\ldots) )
\end{align*}
Note the need to explicitly mark the 9 heartbeats and nest the setting of clock variables 9-deep.
This computation can be described in a \ac{QRE} in a simpler, more concise manner.
Just like the usual regular expressions, simpler \acp{QRE} can be combined into more complex ones. 
We will now use the $\splitop{op}$ combinator (see Fig.~\ref{fig:qre illustrations}): given the input string $w = w_1w_2$ which is a concatenation of strings $w_1$ and $w_2$, and \acp{QRE} $\qre{f},\qre{g}$, $\splitop{op}(\qre{f},\qre{g})$ maps $w$ to the cost value $op(\qre{f}(w_1), \qre{g}(w_2))$, where $op$ is some operator (e.g., averaging).
So let \ac{QRE} $\qre{fourBeats}$ match four consecutive beats in the boolean signal $w$ and let it compute the average cycle length of these 4 beats. Let $inc(x,y)$ be an operation that returns True whenever $0.8x \geq y$.
Then \ac{QRE} $\qre{suddenOnset}$ does the job:
\begin{align*}
\qre{suddenOnset}\,&:=&\,\splitop{inc}(\qre{fourBeats},\,\qre{fourBeats}) 
\\
\qre{fourBeats} &:=& iter_4\!-\!avg(\qre{intervalLength})
\\
\qre{intervalLength} &:=& \splitop{left}(\qre{countzeros}, 1)  \quad /\!/\, left(a,b) \textrm{ returns } a
\end{align*}

The third discriminator takes in a three-vaued signal $w:\Ne \rightarrow \{0,A,V\}$ where a 0 indicates no beat, an $A$ indicates an atrial beat, and a $V$ indicates a ventricular beat.
One \textit{simplified} version of this discriminator detects whether this pattern occurs in the current window: $V 0^{a:b} A 0^{c:d} V 0^{e:f} A 0^{g:h} V$.
Here, $a$ and $b$ are integers, and $0^{a:b}$ indicates between $a$ and $b$ repetitions of 0. 
This can be expressed in discrete-time Metric Temporal Logic~\cite{Koymans90}. 
E.g. the prefix $V 0^{a:b}A$ can be written as
$w=V \implies X ((w=0)\, \until_{[a+1,b]}(w=A))$.
And so on.
This quickly becomes unwieldy as the pattern itself becomes lengthier and with more restrictions on the timing of the repetitions.
On the other hand, this is trivially expressed as a (quantitative) regular expresssion.

Our final example comes from Peak Detection (PD), which takes in a real-valued signal $v:\Ne \rightarrow \Re_{\geq 0}$.
For \textit{one component} of this PD, the objective is to detect when $v(t)$ exceeds a threshold value $h >0 $ which
is reset as a function of the previous peak value.
Thus the logic must remember the value of that peak.  
This necessitates \emph{freeze quantification} of state variables, as used in
Constraint LTL with Freeze Quantification $\CLTLF$~\cite{DEMRI07_ConstraintLTL} ($\freeze_{z=v}$ means that we freeze the variable $z$ to the value of $v$):
\begin{eqnarray*}
	\always (v > h \implies \freeze_{BL=1} \eventually  \left( 
	\formula_{\text{local-max}} \implies  h = 0.8z_2 \right))
	\\
	\formula_{\text{local-max}} \defeq  \freeze_{z_1=v} \nxt 
	(\freeze_{z_2=v} \nxt (z_2 > z_1 \land z_2> v))
\end{eqnarray*}	
The nesting of freeze quantifiers increases the chances of making errors when writing the specification and decreases its legibility.
More generally, monitoring of nested freeze quantifiers complicates the monitors significantly and increases their runtimes. 
E.g., in~\cite{BrimDSV14} the authors show that the
monitoring algorithm for STL with nested freeze 
quantifiers 
is exponential in the number 
of the nested freeze operators in the formula.
This becomes more significant when dealing with the \textit{full} PD, of which the above is one piece.
On the other hand, we have implemented an even more complex PD as a QRE (Section~\ref{sec:qre_wpm}).

The reader will recognize that the operations performed in these tasks are quite common, like averaging, variability, and state-dependent resetting of values, and can conceivably be used in numerous other applications.

This variety of logics required for these tasks, all of which
are fundamental building blocks of ICD operation, means that a temporal logic-based approach to the problem is unlikely to yield a unifying view, whereas QREs clearly do. 
In the rest of the paper, the focus is placed on peak detection, as it is more complicated than discrimination, and offers a strong argument for the versatility and power of \acp{QRE} in medical-device algorithms.
\section{Peaks in the Wavelet Domain}
\label{sec:peaks in wav domain}

Rather than confine ourselves to one particular peak detector, we first describe a general definition of peaks, following the classical work of Mallat and Huang~\cite{MallatH92_Singularity}.
Then two peak detectors based on this definition are presented.
In Section~\ref{sec:experiments}, a third, commercially available, peak detector is also implemented.

\subsection{Wavelet Representations}
\label{sec:wavelets}

This definition operates in the wavelet domain, so a brief overview of wavelets is now provided.
Readers familiar with wavelets may choose to skip this section.
Formally, let $\{\Psi_s\}_{s>0}$ be a family of functions, called \textit{wavelets}, which are obtained by scaling and dilating a so-called \emph{mother wavelet} $\psi(t)$: $\Psi_s(t) = \frac{1}{\sqrt{s}}\psi\left(\frac{t}{s}\right)$.
The \textit{wavelet transform} $\wav$ of signal $\egm: \reals_+ \rightarrow \reals$ is the two-parameter function:
\begin{equation}
\label{eq:cwt}
\wav(s,\,t)=\int\limits_{-\infty}^{+\infty}\,
\egm(\tau)\Psi_s(\tau-t)\,d\tau
\end{equation}
An appropriate choice of $\psi$ for peak detection is the $n^{th}$ derivative of a Gaussian, that is: 
$\psi(t) = \frac{d^n}{dt^n}G_{\mu,\sigma}(t)$.
Eq.~\eqref{eq:cwt} is known as a \emph{Continuous Wavelet Transform} (CWT), and $\wav(s,t)$ is known as the \textit{wavelet coefficient.}

Parameter $s$ in the wavelet $\psi_s$ is known as the \emph{scale} of the analysis.
It can be thought of as the analogue of frequency for Fourier analysis.
A smaller value of $s$ (in particular $s<1$) \emph{compresses} the mother wavelet as can be seen from the definition of $\Psi_s$, so that only values close to $\egm(t)$ influence the value of $\wav(s,t)$
(see Eq.~\eqref{eq:cwt}).
Thus, at smaller scales, the wavelet coefficient $\wav(s,t)$ captures \emph{local} variations of $\egm$ around $t$, and these can be thought of as being the higher-frequency variations, i.e., variations that occur over a small amount of time.
At larger scales (in particular $s > 1$), the mother wavelet is \emph{dilated}, so that $\wav(s,t)$ is affected by values of $\egm$ far from $t$ as well.
Thus, at larger scales, the wavelet coefficient captures variations of $\egm$ over large periods of time.

Fig.~\ref{fig:nsr and its cwt} shows a Normal Sinus Rhythm EGM and its CWT $|\wav(s,t)|$.
The latter plot is known as a \emph{spectrogram}.
Time $t$ runs along the x-axis and scale $s$ runs along the y-axis.
Brighter colors indicate larger values of coefficient magnitudes $|\wav(s,t)|$.
It is possible to see that early in the signal, mid- to low-frequency content is present (bright colors mid- to top of spectrogram), followed by higher-frequency variation (brighter colors at smaller scales), and near the end of the signal, two frequencies are present: mid-range frequencies (the bright colors near the middle of the spectrogram), and very fast, low amplitude oscillations (the light blue near the bottom-right of the spectrogram).

\subsection{Wavelet Characterization of Peaks}
\label{sec:waveletsForPD}
Consider the signal and its CWT 
spectrogram $|\wav(s,t)|$ shown in Fig.~\ref{fig:nsr and its cwt}.
The coefficient magnitude $|\wav(s,t)|$ is a measure of signal power at $(s,t)$.
At larger scales, one obtains an analysis of the low-frequency variations of the signal, which are unlikely to be peaks, as the latter are characterized by a rapid change in signal value.
At smaller scales, one obtains an analysis of high-frequency components of the signal, which will include both peaks and noise.
These remarks can be put on solid mathematical footing \cite[Ch. 6]{waveletMallat}.
\textbf{Therefore, for peak detection one must start by querying CWT coefficients that occur at an appropriately chosen scale $\bar{s}$}.

Given the fixed scale $\bar{s}$, the resulting $|\wav(\bar{s},t)|$ is a function of time.
The next task is to find the \emph{local maxima} of $|\wav(\bar{s},t)|$ as $t$ varies.
The times when local maxima occur are precisely the times when the energy of scale-$\bar{s}$ variations is locally concentrated.
\textbf{Thus peak characterization further requires querying the local maxima at $\bar{s}$}.

Not all maxima are equally interesting; rather, only those with  value above a threshold, since these are indicative of signal variations with large energy concentrated at $\bar{s}$.
\textbf{Therefore, the specification only considers those local maxima with A value above a threshold $\bar{p}$}. 

Maxima in the wavelet spectrogram are not isolated: as shown in~\cite[Thm. 6.6]{waveletMallat}, when the wavelet $\psi$ is the $n^{th}$ derivative of a Gaussian, the maxima belong to connected curves $s\mapsto \gamma(s)$ that are never interrupted as the scale decreases to~0.
These \emph{maxima lines} can be clearly seen in Fig.~\ref{fig:nsr and its cwt} as being the vertical lines of brighter color extending all the way to the bottom.
Multiple maxima lines may converge to the same point $(0,t_c)$ in the spectrogram as $s\rightarrow 0$.
A celebrated result of Mallat and Hwang~\cite{MallatH92_Singularity} shows that \emph{singularities} in the signal always occur at the convergence times $t_c$.
For our purposes, a singularity is a time when the signal undergoes an abrupt change (specifically, the signal is poorly approximated by an $(n+1)^{th}$-degree polynomial at that change-point). 
\textbf{These convergence times are then the peak times that we seek.} 

Although theoretically, the maxima lines are connected, in practice, signal discretization and numerical errors will cause some interruptions. 
Therefore, rather than require that the maxima lines be connected, we only require them to be $(\epsilon,\delta)$-connected.
Given $\epsilon,\delta>0$, an $(\epsilon,\delta)$-\textit{connected curve} $\gamma(s)$ is one such that for any $s$ in its domain, $|s-s'|<\epsilon \implies |\gamma(s)-\gamma(s')| < \delta$.

A succinct description of this \emph{Wavelet Peaks with Maxima} (WPM) is then:
\let\labelitemi\labelitemii
\begin{itemize}
	\item (Characterization \textit{WPM}) Given positive reals $\bar{s}, \bar{p},\epsilon,\delta >0$, a peak is said to occur at time $t_0$ if there exists a $(\epsilon,\delta)$-connected curve $s \mapsto \gamma(s)$ in the $(s,t)$-plane such that $\gamma(0) = t_0$,  $|\wav(s,\gamma(s))|$ is a local maximum along the $t$-axis for every $s$ in $[0,\bar{s}]$, and $|\wav(\bar{s},\gamma(\bar{s}))|\geq \bar{p} $.
\end{itemize}
The choice of values $\bar{s}$, $\epsilon$, $\delta$ and $\bar{p}$ depends on prior knowledge of the class of signals we are interested in.
Such choices are pervasive and unavoidable in signal processing, as they reflect application domain knowledge.
Such a specification is difficult, if not impossible, to express in  temporal and time-frequency logics.
In the next section we show how WPM can be formalized using Quantitative Regular Expressions.

\subsection{Blanking Characterization}
\label{sec:blanking ch}
For comparison, we modify WPM to obtain a peak characterization that is computationally cheaper but suffers some imprecision in peak-detection times.
We call it \emph{Wavelet Peaks with Blanking} (WPB). 
It says that one peak at the most can occur in a time window of size $BL$ samples.
\begin{itemize}
	\item  (Characterization \textit{WPB}) Given positive reals $\bar{s}$, $\bar{p} >0$, a peak is said to occur at time $t_0$ if $|\wav(\bar{s},t_0)|$ is a local maximum along $t$ and $|\wav(\bar{s},t_0)|>\bar{p}$, and there is no peak occurring anywhere in $(t_0, t_0+BL]$.
\end{itemize}
Section~\ref{sec:experiments} compares WPM and WPB on patient electrograms.

\section{A QRE Primer}
\label{sec:qre primer}

\begin{figure*}[t]
	\centering
	\includegraphics[scale=0.3]{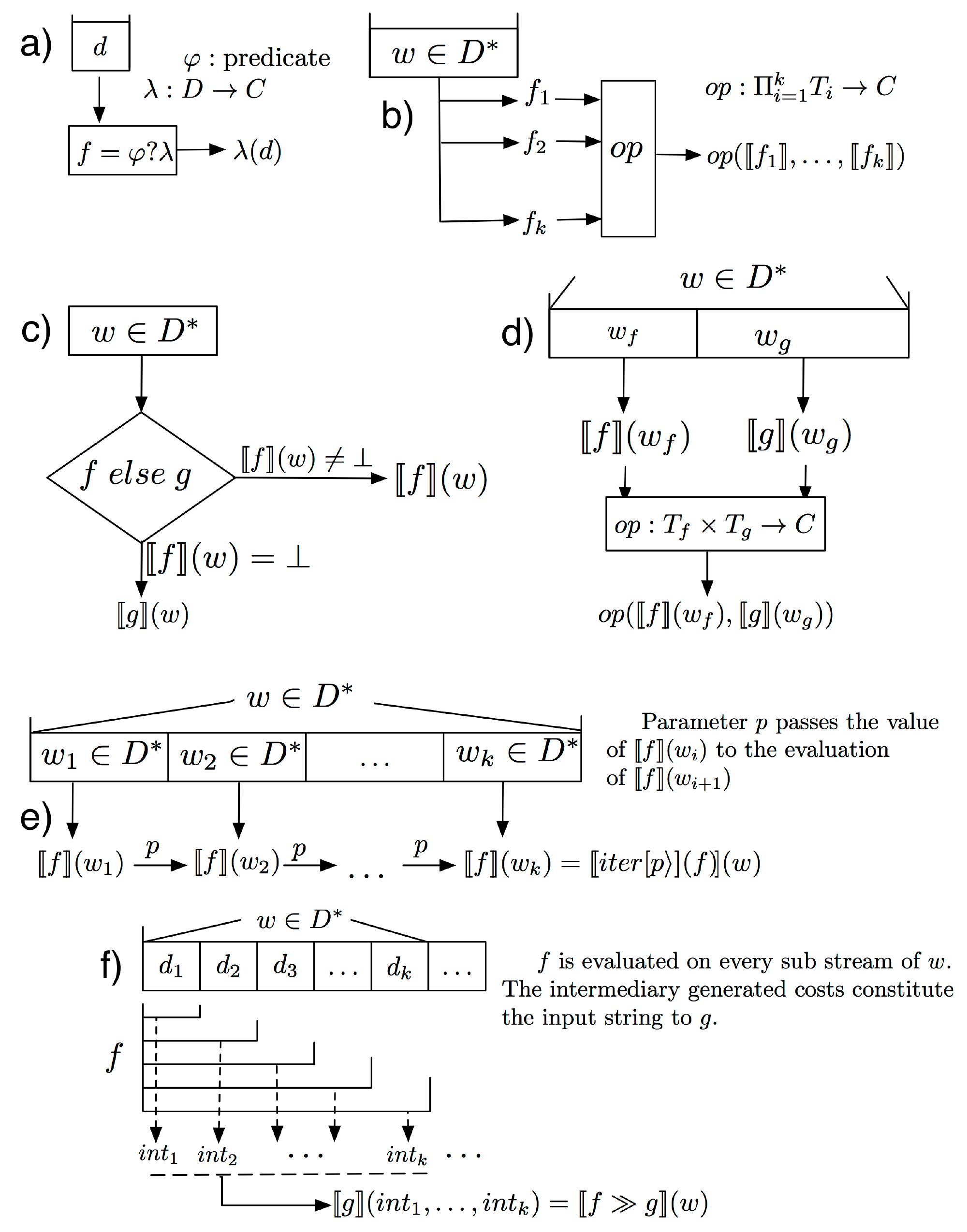}
	\caption{{\small QREs and their combinators. 
			a)~Basic QRE $\varphi?\lambda$ matches one data item $d$ and evaluates to $\lambda(d)$ if $\varphi(d)$ is True. b)~QRE $op(f_1,\ldots,f_k)$ evaluates the $k$ QREs $f_1,\ldots,f_k$ on the same stream $w$ and combines their outputs using operation $op$ (e.g., addition). $f_i$ outputs a value of type $T_i$. 
		c)~QRE $f\; else\; g$ evaluates to $f$ if $f$ matches the input stream; else it evaluates to $g$. 
		d)~QRE $\splitop{op}(f,g)$ splits its input stream in two and evaluates $f$ on the prefix and $g$ on the suffix; the two results are then combined using operation $op$. 
		e)~QRE $iter[p\rangle\!(f)$ iteratively applies $f$ on substreams that match it, analogously to the Kleene-$^*$ operation for REs. Results are passed between iterations using parameter $p$.
		f)~QRE $f \gg g$ feeds the output of QRE $f$ into QRE $g$ as $f$ is being computed. }}
	\label{fig:qre illustrations}
\end{figure*}

An examination of discrimination and PD (Sections~\ref{sec:manylogics} and~\ref{sec:peaks in wav domain}) shows the need for a language that:
1)~Allows a rich set of numerical operations.
2)~Allows  matching of complex patterns in the signal, to select scales and frequencies at which interesting structures exist. 
3)~Supports the synthesis of time- and memory-efficient implementations.
This led to the consideration of \acfp{QRE}. 
A \ac{QRE} is a symbolic regular expression over a data domain $D$, augmented with data
costs from some cost domain $C$.  
A \ac{QRE} views the signal as a
\emph{stream} $w\in D^*$ that comes in one data item at a time. As the Regular Expression (RE) matches
the input stream, the cost of the \ac{QRE} is evaluated.


Formally, consider a set of types $\Tc = \{T_1,T_2,\ldots ,T_k\}$, a data domain $D\,{\in}\,\Tc$,
a cost domain $C\,{\in}\,\Tc$, 
and a parameter set
$X\,{=}\,(x_1,x_2,{\ldots},x_k)$, where each $x_i$ is of type $T_i$.
Then a \ac{QRE} $f$ is a function
\[\val{f}{:}\,D^*\:{\rightarrow}\:({T_1}\,{\times}\,{T_2}\,{\times}\ldots{\times}\,{T_k}\:{\rightarrow}\:{C})\,{\cup}\,\{\bot\}\]
where $\bot$ is the undefined value.
Intuitively, if the input string $w \in D^*$ does not match the RE of $f$, then $\val{f}(w) = \bot$.
Else, $\val{f}(w)$ is a function from ${T_1}\,{\times}\,{T_2}\,{\times}\ldots{\times}\,{T_k}$ to $C$. 
When
a parameter valuation $\bar v \in {T_1}\,{\times}\ldots{\times}\,{T_k}$ is given, this then further evaluates
to a cost value in $C$, namely $\val{f}(w)(\bar v)$. 
Fig.~\ref{fig:qre illustrations} provides an overview of \acp{QRE} and their combinators.

\acp{QRE} can be compiled into efficient \textit{evaluators} that process each
data item in time (or memory) polynomial in the size of the \ac{QRE}
and proportional to the maximum time (or memory) needed to perform an \emph{operation} on a set of cost terms, such as addition, least-squares, etc.
The operations are selected from a set of operations \emph{defined by the user}.
\textit{It is important to be aware that the choice of operations constitutes a trade-off between expressiveness (what can be computed) and complexity (more complicated operations cost more).}
See~\cite{Alur2016} for restrictions placed on the predicates and the symbolic regular expressions.

The declarative nature of \acp{QRE} will be important when writing complex algorithms, without having to explicitly maintain state and low-level data flows.
But as with any new language, \acp{QRE} require some care in their usage. 
Space limitations preclude us from giving the formal definition of \acp{QRE}.
Instead, we will describe what each \ac{QRE} does in the context of peak detection to give the reader a good idea of their ease of use and capabilities.
Fig.~\ref{fig:qre illustrations} illustrates how 
\acp{QRE} are defined and what they compute.
Readers familiar with \acp{QRE} will notice that, when writing the QRE expressions, we occasionally sacrifice strict syntactic correctness for the sake of presentation clarity.

\section{QRE Implementation of Peak Detectors}
\label{sec:qre PD}
We now describe the QREs that implement peak detectors WPM and WPB of Section~\ref{sec:waveletsForPD}.
It is emphasized that even complicated procedures such as these two algorithms can be described in a declarative fashion using \acp{QRE}, without resorting to a programming language or explicitly storing state, etc.

\subsection{QRE for WPM}
\label{sec:qre_wpm}
\begin{figure*}[t]
\centering
\includegraphics[scale=0.5]{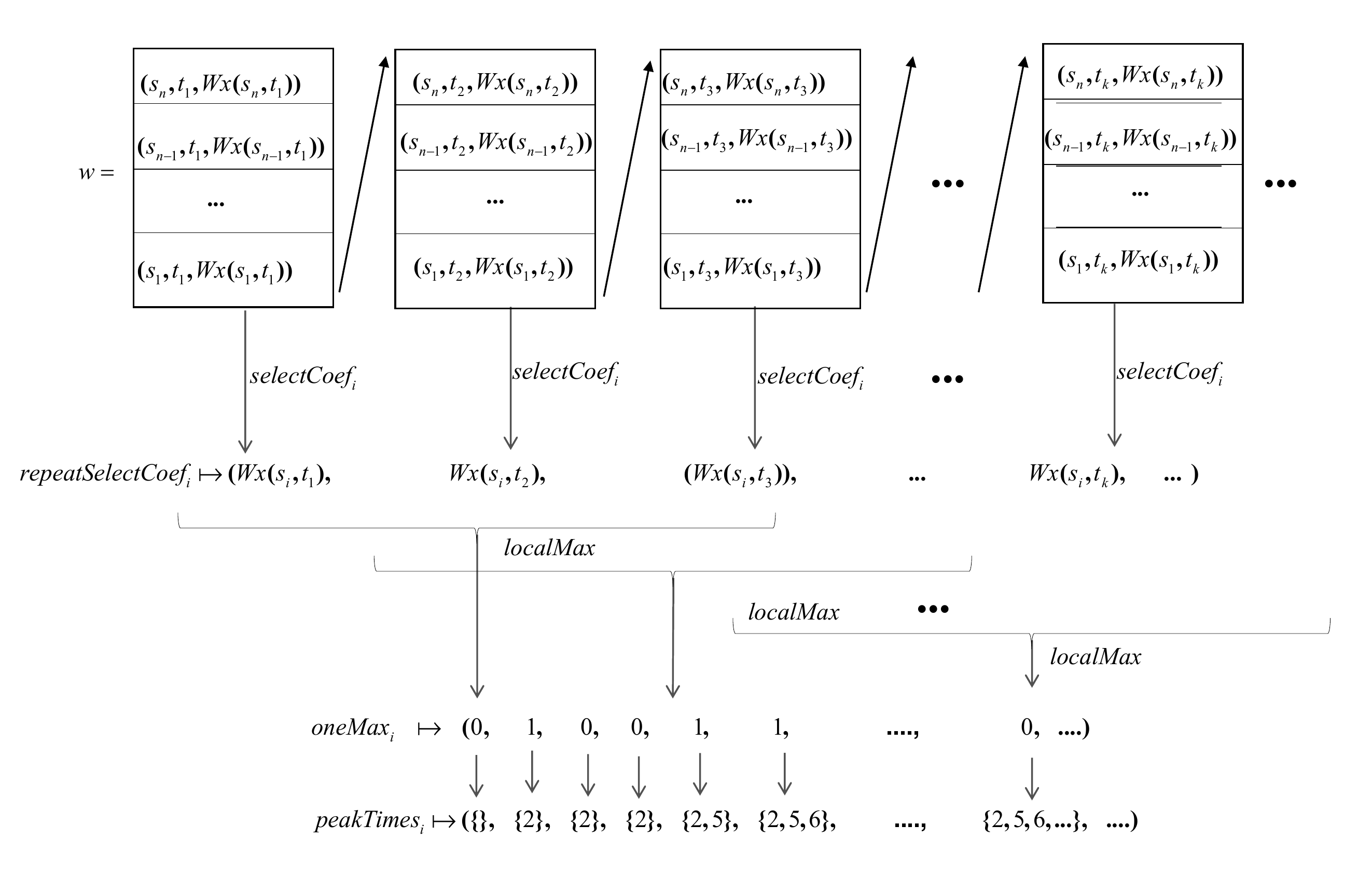}
\caption{QRE $\qre{peakWPM}$}
\label{fig:WPMstream}
\end{figure*}

A numerical implementation of a CWT returns a discrete set of 
coefficients.
Let $s_1 < s_2 <\ldots <s_n$ be the analysis scales and let 
$t_1,t_2,\ldots$ be the signal sampling times.
Recall that a QRE views its input as a stream of incoming data items.
A data item for WPM is 
$d\,=\,(s_i,\,t_j,\,|\wav(s_i,t_j)|) \in D \defeq (\reals_+)^3$.
We use $d.s$ to refer to the first component of $d$, and $d.|\wav(s,t)|$ to refers to its last component.
The input stream $w\in D^*$ is defined by the values from the spectrogram organized in a column-by-column fashion starting 
from 
the highest scale:
\begin{align*}
\label{eq:fullstream}
w = & \underbrace{(s_n,t_1,|\wav(s_n,t_1)|),\ldots,(s_1, t_1,|\wav(s_1,t_1)|)}_{w_{t_1}}
\ldots
\\
& \ldots
\underbrace{(s_n,t_m,|\wav(s_n,t_m)|),\ldots, (s_1,t_m,|\wav(s_1,t_m)|)}_{w_{t_m}}
\end{align*}
Let $s_\sigma$, $1\leq \sigma\leq n$, the the scale that equals $\bar{s}$.
Since the scales $s_i > s_{\sigma}$ are not relevant for peak detection (their frequency is too low), they should be discarded from $w$.
Now, for each scale $s_{i}$, $i \leq \sigma$, we would like to
find those local maxima of 
$|\wav(s_i,\cdot)|$ that are larger than threshold 
$p_i$\footnote{{\small $p_\sigma = \bar{p}$, $p_{i<\sigma} = 0$,
since we 
threshold only the spectrogram values at scale $\bar{s}$.  After this initial thresholding, tracing of maxima lines returns the peaks.}}. 
We build the QRE $\qre{peakWPM}$ bottom-up as follows. 
In what follows, $i= 1,\ldots,\sigma$.
See Fig.~\ref{fig:WPMstream}.
\begin{itemize}
	\item QRE $\qre{selectCoef}_i$ selects the wavelet coefficient magnitude at scale $s_i$ from the incoming spectrogram 
	column $w_t$.
	It must first wait for the entire colum to arrive in a streaming fashion, so it matches $n$ data items (recall there are $n$ items in a column -- see Fig.~\ref{fig:WPMstream}) and returns as cost $d.|\wav(s_i,t)|$.
	\begin{equation*}
	\qre{selectCoef}_i :=\,(d_nd_{n-1}\ldots d_1?\,d_i.|\wav(s_i,t)|).
	\end{equation*}
    \item \ac{QRE} $\qre{repeatSelectCoef}_i$ applies $\qre{selectCoef}_i$ to the latest column $w_t$.
	To do so, it splits its input stream in two: it executes $\qre{selectCoef}_i$ on the last column, and ignores all columns that preceded it using $(d^n)^*$.
    It returns the selected coefficient $|\wav(s_i,t)|$ from the last column.
    \begin{align*}
    \qre{repeatSelectCoef}_{i} := \splitop{right}((d^n)^*,\, 
    \qre{selectCoef}_i)
    \end{align*}
    Combinator $\splitop{right}$ returns the result of operating on the right-hand side of the split, i.e. the suffix.
	\item QRE $\qre{localMax}_i$ matches a string of real numbers of length at least 3:
	$r_1...r_{k-2}r_{k-1}r_k$.
	It returns the value of $r_{k-1}$ if it is larger than $r_k$ and 
	$r_{k-2}$, and is above some pre-defined threshold $p_i$; 
	otherwise, it returns 0. 
	This will be used to detect local maxima in the spectrogram in a moving-window fashion. In detail: 
	\begin{equation}
	\qre{localMax}_i\,\defeq\,\splitop{right}(\reals^*?0,\,\qre{LM}_3)
	\end{equation} 
	$\qre{localMax}_i$ splits the input string in two: the prefix is matched by $\reals^*$ and is ignored.
	The suffix is matched by QRE $\qre{LM}_3$: $\qre{LM}_3$  matches a length-three string and simply returns $1$ if the 
	middle value is a local maximum that is 
	above $p_i$, and returns zero, otherwise.
	\item QRE $\qre{oneMax}_i$ feeds 
	outputs
	of QRE $\qre{repeatSelectCoef}_{i}$ to the \ac{QRE} $\qre{localMax}_i$.
	\[\qre{oneMax}_i \defeq \qre{repeatSelectCoef}_{i} \gg \qre{localMax}_i\]
	Thus, $\qre{oneMax}_i$ ``sees" a string of coefficient magnitudes $|\wav(s_i,t_1)|,|\wav(s_i,t_2)|,\ldots$ generated by (streaming) $\qre{repeatSelectCoef}_i$, and produces a $1$ at the times of local maxima in this string.
	\item QRE $\qre{peakTimes}_i$ collects the 
	times 
	of local maxima at scale $s_i$ into one set.
	\[\qre{peakTimes}_i\,:=\,\qre{oneMax}_i \gg \qre{unionTimes} \]
	It does so by passing the string of 1s and 0s produced by $\qre{oneMax}_i $ to $\qre{unionTimes}$.
	The latter counts the number of 0s separating the 1s and puts that in a set $\Mc_i$.
	Therefore, after $k$ columns $w_t$ have been seen, set $\Mc_i$ 
	contains all 
	local maxima at scale $s_i$ which are above $p_i$ in those $k$ columns.
	\item QRE $\qre{peakWPM}$ is the final QRE.
	It combines results obtained from scales $s_\sigma$ down to $s_1$:
	\begin{equation*}
	\qre{peakWPM} \,:=\, conn_\delta(\qre{peakTimes}_\sigma,...,\,\qre{peakTimes}_1)
	\end{equation*}
	Operator 
	$conn_\delta$\footnote{{\small Operator $conn_\delta$ can be defined recursively as follows:
	$conn_\delta(X, Y)\,=\,\{y \in Y:\exists x\in X:\, |x-y|\leq \delta\},\,  conn_\delta(X_{k},..,X_1)\,=\,conn_\delta(conn_\delta(X_k,..,X_2),\, X_1)$
	}	} checks if the
		local maxima times for each scale (produced by $\qre{peakTimes}_i$) are within a 
		$\delta$ of the maxima at the previous 
	scale.
\end{itemize}
In summary, the complete \ac{QRE} is given top-down by:
\begin{align*}
\qre{peakWPM} &\,:=\, conn_\delta(\qre{peakTimes}_\sigma,...,\,\qre{peakTimes}_1)\\
\qre{peakTimes}_i\,&:= \,\qre{oneMax}_i \gg \qre{unionTimes} \\
\qre{oneMax}_i &\defeq \qre{repeatSelectCoef}_{i} \gg \qre{localMax}_i\\
\qre{localMax}_i\,&:=\, \splitop{right}(\reals^*?0,\,\qre{LM}_3) \\
\qre{repeatSelectCoef}_{i} \,&:=\,
\splitop{right}((d^n)^*,\, 
\qre{selectCoef}_i) \\
\qre{selectCoef}_i\,&:=\,(d_n\ldots d_1?\,d.|\wav(s_i,t)|)
\end{align*}
\subsection{QRE Implementation of WPB}
\label{sec:qre wpb}
\begin{figure}[t]
	\centering
	\includegraphics[scale=0.25]{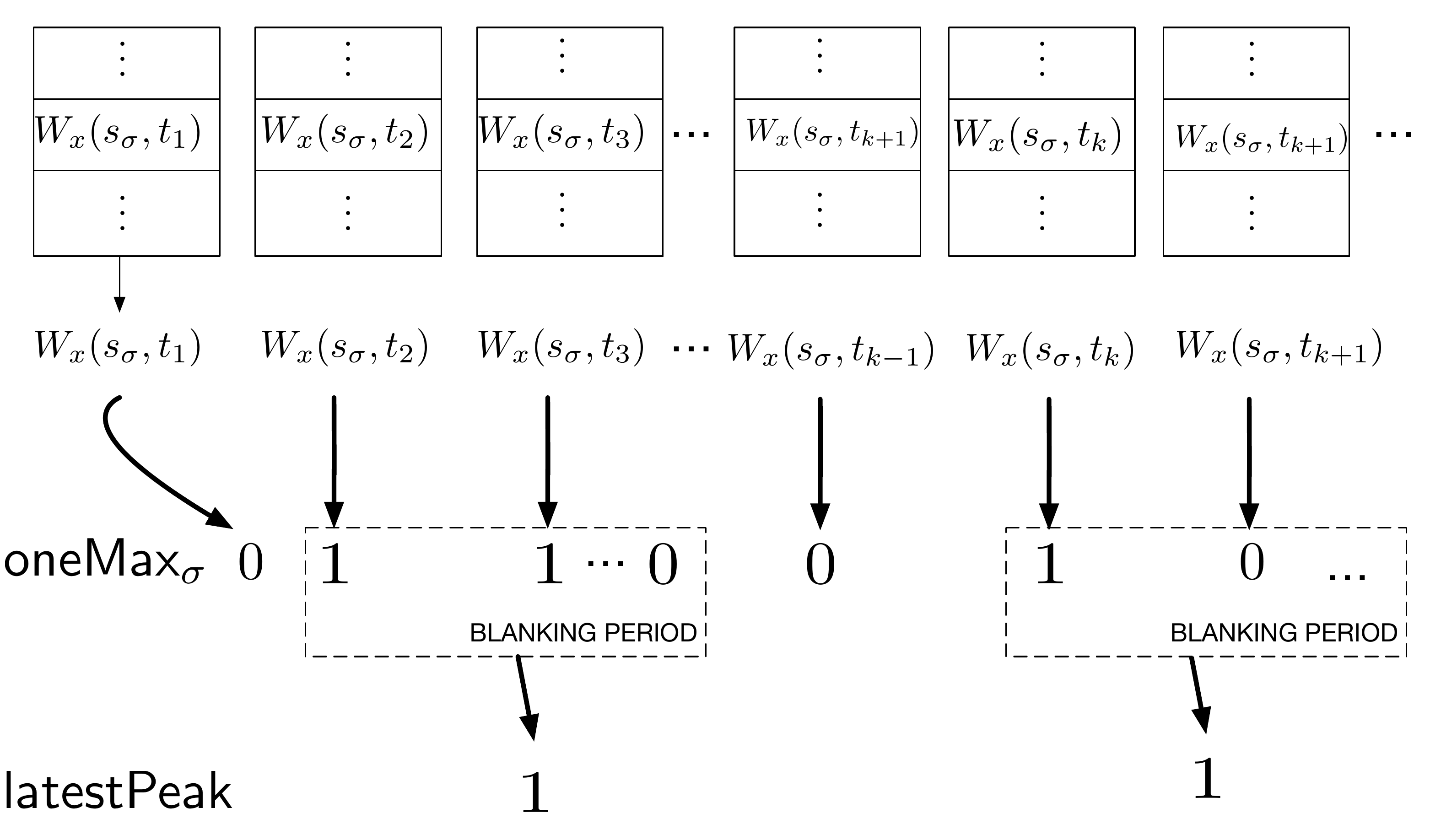}
	\caption{QRE $\qre{peakWPB}$}
	\label{fig:WPB}
\end{figure}

Peak characterization WPB of 
Section~\ref{sec:waveletsForPD} is implemented as \ac{QRE} $\qre{peakWPB}$.
See Fig.~\ref{fig:WPB}.
The input data stream is the same as before.
\begin{itemize}
	\item QRE $\qre{oneMax}_\sigma$ (defined as before) produces a string of 
	$1$s and $0$s, with the 1s indicating local maxima at scale $\bar{s}=s_\sigma$.
	\item QRE $\qre{oneBL}$ matches one blanking duration, starting with the maximum that initiates it.
	Namely, it matches a maximum (indicated by a~1), followed by a blanking period of length $BL$ samples, followed by any-length string without maxima (indicated by $0^*$):
	$\qre{oneBL} \defeq (1\cdot(0|1)^{BL}\cdot0^*)$
	\item QRE $\qre{latestPeak}$ will return a~1 at the time of the latest peak in the input signal:
	$\qre{latestPeak}=\splitop{right}(\qre{oneBL}^*?0,\,1?1)$.
	It does so by matching all the blanking periods up to this point using $\qre{oneBL}^*$ and ignoring them.
	It then matches the maximum (indicated by~1) at the end of the signal.
	\item QRE $\qre{peakWPB}$ feeds the string of 1s and 0s produced by $\qre{oneMax}_\sigma$ to the QRE $\qre{latestPeak}$:
	$\qre{peakWPB}\,=\,\qre{oneMax}_\sigma\gg \qre{latestPeak}$
\end{itemize}

\section{Experimental Results}
\label{sec:experiments}
We show the results of running peak detectors $\qre{peakWPM}$ and $\qre{peakWPB}$ on real patient data, obtained from a dataset of intra-cardiac electrograms.
We also specified a peak detector available in a commercial ICD \cite{icdbook} as \ac{QRE} $\qre{peakMDT}$, and show the results for comparison purposes.
The implmentation uses an early version of the StreamQRE Java library~\cite{MRAIK17}.
Comparing the runtime and memory consumption of different algorithms (including algorithms programmed in QRE) in a consistent and reliable manner requires running a compiled version of the program on a particular hardware platform.
No such compiler is available at the moment, so we don't report such performance numbers.

The results in this section should not be interpreted as definitively establishing the superiority of one peak detector over another, as this is not this paper's objective.
Rather, the objective is to highlight the challenges involved in peak detection for cardiac signals, an essential signal-processing task in many medical devices.
In particular, by highlighting how different detectors perform on different signals, it establishes the need for a formal (and empirical) understanding of their operation on classes of arrhythmias.
This prompts the adoption of a formal description of peak detectors for further joint analysis with discrimination.

\begin{figure*}[t!]
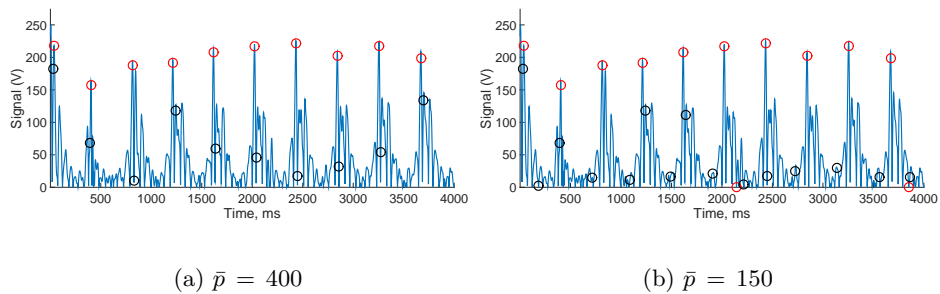

	\centering
	\begin{subfigure}[t]{0.5\textwidth}
		\centering
		\includegraphics[height=1.2in]{figures/A176WPBnWAVs80thr400}
		\label{fig:VT WPM WPB p400}
		\caption{$\bar{p}\,=\,400$}
	\end{subfigure}%
	~ 
	\begin{subfigure}[t]{0.5\textwidth}
		\centering
		\includegraphics[height=1.2in]{figures/A176WPBnWAVs80thr150}	
		\label{fig:VT WPM WPB p150}
		\caption{$\bar{p}\,=\,150$}
	\end{subfigure}
	\caption{\small $\qre{peakWPM}$-detected peaks (red circles) and $\qre{peakWPB}$-detected peaks (black circles) on a VT rhythm.}
	\label{fig:VT WPM WPB}
\end{figure*}


Fig.~\ref{fig:VT WPM WPB} presents one rectified EGM signal of a \acf{VT} recorded from a patient. 
Circles (indicating detected time of peak) show the result of 
running $\qre{peakWPM}$ (red circles) and $\qre{peakWPB}$ (black circles). 
These results were obtained for $\bar{s} = 80$, $BL = 150$, and different 
values of $\bar{p}$.
The first setting of $\bar{p}$ (Fig.~\ref{fig:VT WPM WPB}~(a)) for both \acp{QRE} was chosen to yield the best performance.
This is akin to the way cardiologists set the parameters of commercial \acp{ICD}: they observe the signal, then set the parameters.
We refer to this as the \textit{nominal setting}.
Ground-truth is obtained by having a cardiologist examine the signal and annotate the true peaks.

We first observe that the peaks detected by $\qre{peakWPM}$ match the ground-truth; i.e., the nominal performance of $\qre{peakWPM}$ yields perfect detection.
This is not the case with $\qre{peakWPB}$.
Next, one can notice that the time precision of detected peaks with $\qre{peakWPM}$ is
higher than with $\qre{peakWPB}$ due to maxima lines tracing down to the zero scale. 
Note also that the results of \qre{peakWPM} are stable for various parameters settings. 
Improper thresholds $\bar{p}$ or scales $\bar{s}$ degrade 
the results only slightly (compare locations of red circles on 
Fig.~\ref{fig:VT WPM WPB} (a) with 
Fig.~\ref{fig:VT WPM WPB} (b)). 
By contrast, $\qre{peakWPB}$ detects additional false peaks 
(compare black circles in Figs.~\ref{fig:VT WPM WPB} (a) and (b)). 
\begin{figure*}[t!]
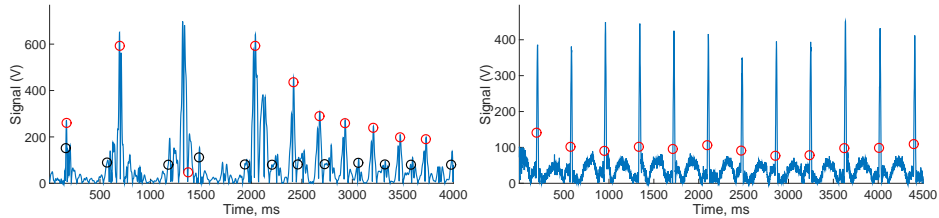

	\centering
	\begin{subfigure}[t]{0.5\textwidth}
		\centering
		\includegraphics[height=1.2in]{figures/A173WAVnWDTs100thr300}
	\end{subfigure}%
	~ 
	\begin{subfigure}[t]{0.5\textwidth}
		\centering
		\includegraphics[height=1.2in]{figures/A181WDTsnominal}	
	\end{subfigure}
	\caption{\small WPM and $\qre{peakMDT}$ running on a VF rhythm (left) and $\qre{peakMDT}$ running on an NSR rhythm (right).}
	\label{fig:VF WPM MDT}
\end{figure*}

Fig.~\ref{fig:VF WPM MDT} (left)
shows WPM (red circles) running on a Ventricular Fibrillation (VF) rhythm, which is a potentially fatal disorganized rhythm.
Again, we note that WPM finds the peaks.

Detector MDT works almost perfectly with nominal parameters settings on 
any Normal Sinus Rhythm (NSR) signal (see Fig.~\ref{fig:VF WPM MDT} right). 
NSR is the ``normal" heart rhythm.
The detected peak times are slightly early because $\qre{peakMDT}$ declares a peak when the signal exceeds a time-varying threshold, rather than when it reaches its maximum.
Using the same 
nominal parameters on more disorganized EGM signals with higher 
variability in amplitude, such as VF, does not produce proper 
results; see the black circles in Fig.~\ref{fig:VF WPM MDT} left.


\section{Related Work}
\label{sec:relatedWork}
Signal Temporal Logic (STL)~\cite{Maler2004} was designed for the specification of temporal, real-time properties over real-valued signals and has been used in many applications including the differentiation of medical signals~\cite{stl-learning,Bartocci2014}. 
In~\cite{BrimDSV14}, STL was augmented with a signal value \emph{freeze operator} that 
 allows one to express oscillatory patterns,
but it is not possible to use it to discriminate
oscillations within a particular frequency range.  
The spectrogram of a signal can be represented as a 2D map (from time and scale to amplitude) and
one may think to employ a spatial-temporal logic such as SpaTeL~\cite{Haghighi2015} or Signal Spatio-Temporal Logic (SSTL)~\cite{Nenzi2015} on spectrograms.
However, both of their underlying spatial models, graph structures for SSTL and quadtrees for SpaTeL, are not appropriate for this purpose.
Logics for describing frequency and temporal properties have been proposed, including
Time-Frequency Logic (TFL) in~\cite{Donze2012} and the approach in~\cite{ChakarovFS2012_TFL}.
TFL is not sufficiently expressive for peak detection 
because it lacks the necessary mechanisms to quantify over variables or to freeze 
their values. 
Timed regular expressions~\cite{Asarin02_TRE,Ulus16_Montre,UlusFAM14} 
extend regular expressions by clocks and are expressively equivalent 
to timed automata, but cannot express the computations required for 
the tasks covered in this paper. Even the recent work proposed 
in~\cite{FerrereMNU15} on measuring signals with timed patterns is not 
of help in our application, since it does not handle, neither in the specification nor 
in the measurement, the notion of local minima/maxima that is necessary
for peak detection. Furthermore, the operator of measure is separated 
by the specification of the pattern to match.

SRV~\cite{DAngelo2005} is a stream runtime \textit{verification} language that requires explicit encoding of relations between input and output streams, which is an awkward way of encoding the complex tasks of this paper.
Moreover, unlike Boolean SRVs~\cite{Bozzelli2014}, \acp{QRE} allow multiple unrestricted data types in intermediary computations and a number of their questions are decidable for these arbitrary types.

\section{Conclusions and Future Work}
\label{sec:future}

The tasks of discrimination and peak detection, fundamental to arrhythmia-discrimination algorithms, are easily and succinctly expressible in \acp{QRE}. 
One obvious limitation of \acp{QRE} is that they only allow regular matching, though this is somewhat mitigated by the ability to chain QREs (though the streaming combinator $\gg$) to achieve more complex tasks.
One advantage of programming in \acp{QRE} is that it automatically provides us with a base implementation, whose time and memory complexity is independent of the stream length.

As future work, it will be interesting to compile a QRE into C or assembly code to measure and compare actual performance on a given hardware platform.
Also, just like an RE has an equivalent machine model (DFA), 
a \ac{QRE} has an equivalent machine 
model in terms of a deterministic finite-state transducer~\cite{Alur2016}.
This points to an analysis of a 
QRE's correctness and efficiency beyond testing.
Two lines of inquiry along these lines are promising in the 
context of medical devices.

\textbf{Probabilistic analysis.}
Assume a probabilistic model of the \ac{QRE}'s input strings.
For medical devices, such a model might be learned from data.
We may then perform a statistical analysis of the output of the \ac{QRE} under such an input model. 
In particular, we may estimate how long it takes the ICD to detect a fatal arrhythmia, or the probability of an incorrect detection by the ICD.

\textbf{Energy calculations.}
We may compute the energy consumption of an algorithm that is expressed as a QRE, by viewing consumption as another quantity computed by the QRE.
Alternatively, we may label the transitions of the underlying DFA by ``energy terms", and levarage analysis techniques of weighted automata to analyze the energy consumption.
Energy considerations are crucial to implanted medical devices that must rely on a battery, and which require surgery to replace a depleted battery.

\section*{Acknowledgments}
The authors would like to thank Konstantinos Mamouras for insightful discussions about QREs and for providing the QRE Java library we used in this paper.  This work is supported in part by %
AFOSR Grant FA9550-14-1-0261 
and NSF Grants
IIS-1447549, 
CNS-1446832, 
CNS-1446664, 
CNS-1445770, 
and CNS-1445770. 
E.B.\ and R.G.\ acknowledge the partial support of the Austrian National Research Network  S 11405-N23 and S 11412-N23 (RiSE/SHiNE) of the Austrian Science Fund (FWF) and the ICT COST Action IC1402 Runtime Verification beyond Monitoring (ARVI).

The authors would like to thank Konstantinos Mamouras for insightful discussions about QREs and for providing the QRE Java library we used in this paper.  This work is supported in part by 
AFOSR Grant FA9550-14-1-0261 
and NSF Grants
IIS-1447549,
CNS-1446832,
CNS-1446664,
CNS-1445770,
and CNS-1445770.
E.B.\ and R.G.\ acknowledge the partial support of the Austrian National Research Network  S 11405-N23 and S 11412-N23 (RiSE/SHiNE) of the Austrian Science Fund (FWF) and the ICT COST Action IC1402 Runtime Verification beyond Monitoring (ARVI).

The authors would like to thank Konstantinos Mamouras for insightful discussions about QREs and for providing the QRE Java library we used in this paper. This work is supported in part by %
AFOSR Grant FA9550-14-1-0261
and NSF Grants
IIS-1447549,
CNS-1446832, 
CNS-1446664, 
CNS-1445770, 
and CNS-1445770, 
and Austrian National Research Network grants 
S 11405-N23 and S 11412-N23 (RiSE/SHiNE of the FWF) and the ICT COST Action IC1402 Runtime Verification beyond Monitoring.
\bibliographystyle{abbrv}
\bibliography{chapters/sigprocHSCC17}
\end{document}